\documentstyle[preprint,aps]{revtex}
\begin{document}
\input{epsf}
\draft
\newfont{\form}{cmss10}
\newcommand{\e}{\varepsilon} 
\renewcommand{\b}{\beta} 
\newcommand{\unity}{1\kern-.65mm \mbox{\form l}}
\newcommand{\D}{D \raise0.5mm\hbox{\kern-2.0mm /}}
\newcommand{\A}{A \raise0.5mm\hbox{\kern-1.8mm /}}
\def\pmb#1{\leavevmode\setbox0=\hbox{$#1$}\kern-.025em\copy0\kern-\wd0
\kern-.05em\copy0\kern-\wd0\kern-.025em\raise.0433em\box0}
\def\D{\hbox{\hbox{${D}$}}\kern-1.9mm{\hbox{${/}$}}}
\def\kbar{\hbox{$k$}\kern-0.2truecm\hbox{$/$}}
\def\nbar{\hbox{$n$}\kern-0.23truecm\hbox{$/$}}
\def\pbar{\hbox{$p$}\kern-0.18truecm\hbox{$/$}}
\def\nhbar{\hbox{$\hat n$}\kern-0.23truecm\hbox{$/$}}
\newcommand{\dif}{\hspace{-1mm}{\rm d}}
\newcommand{\dil}[1]{{\rm Li}_2\left(#1\right)}
\newcommand{\diff}{{\rm d}} 

\title{Infrared singularities in the null-plane bound-state equation
when going to 1+1 dimensions}
\author{A. Bassetto ($^*$)}
\address{CERN, Theory Division, CH-1211 Geneva 23, Switzerland\\
INFN, Sezione di Padova, Padua, Italy}
\maketitle
\begin{abstract}
In this paper we first consider the null-plane bound-state equation 
for a $q \bar q$ pair in 1+3 dimensions and in the lowest-order 
Tamm-Dancoff approximation. Light-cone gauge is chosen with a causal 
prescription for
the gauge pole in the propagator. Then we show that this equation,
when dimensionally reduced to 1+1 dimensions, becomes 't Hooft's 
bound-state equation, which is characterized by an ${\rm x}^+$-
instantaneous interaction. The deep reasons for this coincidence 
are carefully discussed.
\end{abstract} 

\vskip 2.0truecm
PACS numbers: 11.10.Kk, 11.10St, 12.38.Bx

Keywords: Null-plane perturbation theory, bound-state equation, 
lower dimensions.
\vskip 3.0truecm
\noindent
($^*$) Present address:
Dipartimento di Fisica ``G. Galilei", Via Marzolo 8, 35131
Padua, Italy.
\vfill\eject

\narrowtext

\section{Introduction}
\noindent
One of the challenging problems confronting gauge theories is the
transition from theories defined in the usual 1+3 dimensions to
1+1 dimensional theories. In turn 1+1 dimensional theories are
interesting as sometimes they are solvable, or, at least, they
provide useful insights in non-perturbative phenomena.

A central role is played by the choice of the light-cone
gauge, owing to its natural partonic interpretation. On the other hand
this gauge, at least in perturbative treatments, exhibits more severe
infrared (IR) singularities. 

Yang-Mills theories in light-cone gauge were first quantized on a null
plane (light-front quantization \cite{KOGUT}). In this procedure
the gauge pole in the polarization tensor occurring in the free propagator
is treated according to the Cauchy principal value (CPV) prescription,
which has the merit of being ``real'', namely not to contribute to
the propagator absorptive part. However, in so doing, a conflict
is induced with the usual ``Feynman'' pole, which, on physical grounds,
in 1+3 dimensions
must be prescribed in a causal way. This conflict can for instance be
seen as the occurrence of extra unwanted terms when perturbative integrals
undergo a Wick rotation \cite{BASNA}.

To remedy this situation, a causal prescription (ML) was proposed in
refs.\cite{MANDEL},\cite{LEIBB} for the gauge pole; this prescription
was in turn derived by equal-time canonical quantization in ref.\cite{BASSE}
and shown to be mandatory in 1+3 dimensions for a consistent renormalization
\cite{BASNA}.

When ${\rm x}^+$-ordered perturbation theory is used, 
more severe IR singularities occur, which often have been regularized by 
means of artificial cutoffs. On the other hand, the ML prescription 
cannot be easily implemented.
This difficulty is carefully explained in ref.\cite{SOPER93} in which
the bound-state equation for a $q \bar q$ pair is considered 
in the lowest-order Tamm-Dancoff approximation \cite{TAMM45}.
The relevance 
of using a causal prescription for the gauge pole is fully recognized
and a concrete solution for implementing the ML prescription 
is proposed.

The situation drastically changes in 1+1 dimensions.
Here  ultraviolet (UV) singularities no longer occur, hence there
is no need of renormalization.
Both equal-time and null-plane quantization seem a priori viable
\cite{BASNAR}.  The latter indeed does no longer conflict with causality as
no vector degrees of freedom propagate, the gauge
field only providing an ``instantaneous'' potential between fermions.
Canonical quantization suggests the CPV prescription on (both) Feynman
and gauge poles.
\footnote{Of course
``Feynman'' and gauge pole have to be treated coherently; we remind the reader
that the product ${1\over {[q^+]}_{CPV}}{1\over {[q^+]}_{ML}}$ does not
define a distribution.}

A celebrated example of this theory in the large-$N$ approximation
is 't Hooft's bound-state equation
\cite{THOOFT}. From it a beautiful physical
picture emerges with meson bound states lying on rising Regge
trajectories. The counterpart of this equation in equal-time
quantization was proposed by Wu \cite{WUEQU}, a quite difficult two-variable 
integral equation, whose (approximate) solution for particular values of
external parameters has been obtained only very recently \cite{BASHU}. 
The resulting physical picture
is quite different from 't Hooft's; in particular no rising Regge 
trajectories are found.

On the other hand, if the 1+1 theory is to be considered as the limit
of a theory in higher dimensions, then the equal-time formulation
(with related causal prescription) seems unavoidable. This is also the
conclusion one reaches when considering a perturbative Wilson loop
calculation at ${\cal O}(g^4)$ \cite{BASGRI}: Feynman and light-cone
gauges provide the same result, even in the limit $d\to 2$, 
only when canonically quantized at equal time.
This result in turn is quite different from the one derived using the
instantaneous potential coming from null-plane quantization. 

Two different theories thus seem to exist in 1+1 dimensions, one being 
the limit of theories in higher dimensions, the other being simpler and
endowed with nice physical consequences. We would like to stress that the 
difference is not in technical details: the two formulations have a
different content of degrees of freedom \cite{BASNA}.

Still, we show that the bound-state equation, in the lowest order Tamm-Dancoff
approximation and with a causal prescription on the gauge pole, 
when dimensionally reduced
to 1+1 dimensions, coincides with 't Hooft's 
equation, in spite of the fact that interaction is here described 
by an ${\rm x}^+$ instantaneous potential.
As a consequence, in this particular instance, the prescription on the poles
turns out to be irrelevant.
This phenomenon is rooted in the cancellation of IR singularities 
between ``real'' and ``virtual'' contributions
\cite{BASMORI}.

The above considerations motivate the present work. We start from
the concrete lowest-order Tamm-Dancoff approximation of ref.\cite{SOPER93}
in 1+3 dimensions (Sect. 2) and then, in Sect. 3, dimensionally reduce 
it to the 1+1 dimensional case: starting from the
``causal'' formulation of the bound-state equation, we show
that it eventually coincides with the one in which the interaction
is mediated by an ${\rm x}^+$-instantaneous potential, namely
with 't Hooft's equation, in spite
of the seemingly different physical inputs. The reason for this coincidence 
as well as further considerations are drawn in Sect. 4.

\smallskip

\section{The Bethe-Salpeter equation in 1+3 dimensions} 
\noindent
In this section we recall concepts and results developed in ref. \cite{SOPER93},
which the reader is invited to consult. We follow the notation
used there.

The integral equation for a bound state in the $q \bar q$ channel is 
considered in the null-plane formulation, $x^+\equiv (x^0+x^3)/\sqrt 2$
playing the role of time. The idea behind this framework is
that partons cannot pop up spontaneously from the vacuum, 
when the theory is quantized in a ``physical''
gauge; one usually chooses the light-cone gauge $A_- \equiv (A_0-A_3)/\sqrt 2
=0.$
Then a truncation on the number of partons allowed in the wave
function (Tamm-Dancoff approximation \cite{TAMM45}) becomes viable.

For a deeper insight as well as physical motivations, the reader should 
consult the abundant literature on the subject (see references in 
\cite{SOPER93}).

In light-front calculations, singularities occur in the 
IR region of $p^+$, which require a suitable prescription to be handled.
The situation becomes worse in the gauge $A_-=0$, as gauge-dependent 
singularities conspire with the previous ``Feynman'' ones and must be treated 
together in a consistent way.

We consider a meson with momentum
\begin{equation}
P^{\mu} =\Big(P^+,{{{\bf P}^2+M^2}\over {2P^+}},{\bf P}\Big),
\label{momentum}
\end{equation}
which is composed of a quark and an antiquark. The meson state vector
is normalized by
\begin{equation}
\langle P^+,{\bf P}|\hat P^+,\hat {\bf P}\rangle =(2\pi)^3 2P^+ \delta(P^+-
\hat P^+)\delta ({\bf P}- \hat {\bf P}). 
\label{norma}
\end{equation}
Next we consider the Bethe-Salpeter wave function for the meson at ${\bf P}=0$
\begin{equation}
\Phi(p)_{\alpha \beta}= \int d^4{\rm x} e^{ip{\rm x}} \langle0|\psi_{\alpha}
({\rm x})\bar 
\psi_{\beta}(0)|P^+,{\bf 0}\rangle,
\label{bethe}
\end{equation}
where $\psi_{\alpha}$ is the quark field.
From eq. (\ref{bethe}) one can project the null-plane wave function $\psi$
\cite{SOPER93}
\begin{equation}
\psi(x,{\bf p};s_1,s_2)= {1 \over {2P^+}} \int {dp^- \over {2\pi}}
\bar u(xP^+,{\bf p};s_1) \gamma^+ \Phi(p)\gamma^+ v((1-x)P^+,-{\bf p};s_2),
\label{nullwave}
\end{equation}
where $x=p^+/P^+$,
normalized as
\begin{equation}
1 = (2\pi)^{-3} \int_0^1 {dx \over {2x(1-x)}} \int d{\bf p} \sum_{s s'}
|\psi(x,{\bf p};s,s')|^2.
\label{norma1}
\end{equation}
The spinors $u$ and $v$ are normalized to
\begin{equation}
\bar u_{\alpha}(p^+,{\bf p};s)\gamma^+u_{\alpha}(p^+,{\bf p};s')=
\bar v_{\alpha}(p^+,{\bf p};s)\gamma^+v_{\alpha}(p^+,{\bf p};s')=
2p^+\delta_{s s'}.
\label{spinors}
\end{equation}
If we denote by $S(p)=[-i(p\cdot \gamma-m)]^{-1}$ 
the free fermion propagator and by $\Sigma (p)$
the fermion self-energy, the Bethe-Salpeter equation takes the form
\begin{eqnarray}
\label{Bsalpeter}
\Phi(p)_{\alpha \beta}&=&S(p)_{\alpha \alpha'} S(p-P)_{\beta' \beta}
\int {d^4k\over {(2\pi)^4}} T(p,k)_{\alpha' \alpha'' \beta' \beta''} 
\Phi(k)_{\alpha'' \beta''} \nonumber \\
&+& S(p)_{\alpha \alpha'}[-i\Sigma (p)_{\alpha' \alpha''}]
\Phi(p)_{\alpha'' \beta}
+\Phi(p)_{\alpha \beta''}[-i\Sigma (p-P)_{\beta'' \beta'}]S(p-P)_{\beta' \beta}
\nonumber \\
&-&S(p)_{\alpha \alpha'}[-i\Sigma (p)_{\alpha' \alpha''}]
\Phi(p)_{\alpha'' \beta''} [-i\Sigma (p-P)_{\beta'' \beta'}]
S(p-P)_{\beta' \beta},
\end{eqnarray} 
where summation over repeated indices is understood.

In the first term, $T(p,k)$ represents the Bethe-Salpeter kernel, consisting 
of all two-particle irreducible diagrams. We shall consider for it the first
perturbative approximation, namely one-gluon exchange. Similarly the
self-energy will be replaced by its one-loop 
approximation $\Sigma_1$. Renormalization in the $\overline{{\rm MS}}$
scheme is 
understood. 

\subsection{The ``real'' diagram contribution}
\noindent
Let us begin by considering the first term in eq. (\ref{Bsalpeter}),
in the approximation
we have just mentioned:
\begin{eqnarray}
\label{realterm}
\Phi(p)_{1GE}&=&i C_F g^2 S(p)
\int {d^4k\over {(2\pi)^4}} \gamma_{\alpha}\Phi(k) \gamma_{\beta} 
\nonumber \\
&\times& {N^{\alpha \beta}(k-p)\over {[(k-p)\cdot n][(k-p)^2+i\epsilon]}}S(p-P),
\end{eqnarray} 
$C_F$ being the Casimir constant of the fundamental representation and
$$N^{\alpha \beta}(q)=-q\cdot n \, g^{\alpha \beta}+ q^{\alpha}n^{\beta}+
q^{\beta}n^{\alpha}$$ the numerator of the gluon propagator. The gauge fixing 
null vector
$n^{\mu}=(0,1,0,0), n\cdot A=0$ appears also in the denominator and gives
rise to the mentioned gauge dependent singularity at $(k-p)\cdot n=0$.

A simple algebra now gives
\begin{eqnarray}
\label{psirealterm}
\psi_{1GE}(x,{\bf p};s_1,s_2)&=&-i C_F g^2 {{x(1-x)P^+}\over {\pi}}
{{\bar u(xP^+,{\bf p};s_1)}\over {p^2-m^2+i\epsilon}}
\int dp^-
\int {d^4k\over {(2\pi)^4}} \gamma_{\alpha}\Phi(k) \gamma_{\beta} 
\nonumber \\
&\times& {N^{\alpha \beta}(k-p)\over {[(k-p)\cdot n][(k-p)^2+i\epsilon]}}
{{v((1-x)P^+,-{\bf p};s_2)}\over {(P-p)^2-m^2+i\epsilon}}.
\end{eqnarray} 

In ref.\cite{SOPER93} it is carefully explained how the Tamm-Dancoff
approximation allows the above quantity $\psi_{1GE}$ to be expressed
in terms of the null-plane wave function $\psi$. We are not going to repeat
the argument and simply quote the result:
\begin{eqnarray}
\label{psirealapp}
&&\psi_{1GE}(x,{\bf p};s_1,s_2)=-\int {{d^2{\bf k}}\over {(2\pi )^3}}
\int dy \int {{dk^-}\over {2\pi}} \int {{dp^-}\over {2\pi}} 
 F(x,y,k^--p^-,{\bf k}, {\bf p})
\nonumber \\
&\times& \Big[{(k^--\omega(y,{\bf k}^2)+i\epsilon \ {\rm sign}(y))}^{-1}-
{(k^--{\cal E}+\omega(1-y,{\bf k}^2)-i\epsilon\ {\rm sign}(1-y))}^{-1}\Big]
\nonumber \\
&\times& \Big[{(p^--\omega(x,{\bf p}^2)+i\epsilon\ {\rm sign}(x))}^{-1}-
{(p^--{\cal E}+\omega(1-x,{\bf p}^2)-i\epsilon\ {\rm sign}(1-x))}^{-1}\Big]
\nonumber \\
&\times&{\Big[P^+(y-x)\Big]}^{-1}{\Big[2P^+(y-x)(k^--p^-)-
({\bf k}-{\bf p})^2 +i\epsilon\Big]}^{-1},
\end{eqnarray} 
where $y={{k^+}\over {P^+}}$, 
$\omega(y,{\bf k}^2)={{{\bf k}^2+m^2} \over {2yP^+}}$,
${\cal E}={{M^2}\over {2P^+}}$ and $F$ is a short-hand notation for the
quantity
\begin{eqnarray}
\label{bigeffe}
&&F(x,y,k^--p^-,{\bf k}, {\bf p})=-{{C_F g^2}\over {4P^+y(1-y)}}
\sum_{s_1's_2'}\psi(y,{\bf k};s_1',s_2')\bar u(xP^+,{\bf p};s_1)
\gamma_{\alpha}u(yP^+,{\bf k};s_1') \nonumber \\
&&{{N^{\alpha \beta}(k-p)}\over {{\cal E}-\omega(1-x,{\bf p}^2)-
\omega(x,{\bf p}^2)+i\epsilon}}\bar v((1-y)P^+,-{\bf k};s_2')\gamma_{\beta}
v((1-x)P^+,-{\bf p};s_2).
\end{eqnarray} 
Null-plane perturbation theory is recovered by
performing the integrations over $k^-$ and $p^-$. The function $F$
depends on them linearly; therefore these integrations would be
simple, were the gauge singularity $P^+(y-x)=0$ in the denominator
prescribed in such a way as not to involve minus components. 
Then contour
integrations would lead to the result \cite{SOPER93}
\begin{eqnarray}
\label{psiwrong}
&&\psi_{1GE}(x,{\bf p};s_1,s_2)=\theta(1-x) \theta(x)
\int {{d^2{\bf k}}\over {(2\pi )^3}}
\nonumber \\
&\times&\Big\{\int_{x}^{1} dy {\Big[P^+(y-x)\Big]}^{-1}
 {{F(x,y,{\cal E}-\omega(1-y,{\bf k}^2)-\omega(x,{\bf p}^2),
{\bf k}, {\bf p})}\over {2(y-x)P^+[{\cal E}-\omega(1-y,{\bf k}^2)
-\omega(x,{\bf p}^2)]-({\bf k}-{\bf p})^2+i\epsilon}}
\nonumber \\
&+&\int_{0}^{x} dy {\Big[P^+(y-x)\Big]}^{-1}
 {{F(x,y,-{\cal E}+\omega(y,{\bf k}^2)+\omega(1-x,{\bf p}^2),
{\bf k}, {\bf p})}\over {2(x-y)P^+[{\cal E}-\omega(y,{\bf k}^2)
-\omega(1-x,{\bf p}^2)]-({\bf k}-{\bf p})^2+i\epsilon}}\Big\}, 
\end{eqnarray}
where the support of the function $\psi$ has been explicitly
exhibited. 
 
Unfortunately the above expression is meaningless as there are 
manifest singularities 
at the extrema of integration. In other words the gauge singularity 
calls for a prescription before integrating over the minus components.

In refs.\cite{MANDEL}, \cite{LEIBB} and \cite{BASSE},
arguments are presented in favour of the {\it causal} prescription (ML)
\begin{equation}
\label{mandel}
{1\over {[q^+]_{ML}}}
\equiv {1\over {q^++i\epsilon \ {\rm sign}(q^-)}} =
{{q^-}\over {q^+q^-+i\epsilon}},
\end{equation} 
which would not conflict with the (causal) ``Feynman'' poles, 
allowing for a Wick's rotation without extra contributions. This would
not be the case for the CPV prescription,
suggested in \cite{KOGUT}.

A problem then arises in eq. (\ref{psirealapp}), as the integrations over 
the minus components can no longer be done in a straightforward way.

The solution proposed in ref.\cite{SOPER93} is to perform a subtraction,
using the identity (see also ref.\cite{BASSET})
\begin{equation}
\label{identity}
\int_{-\infty}^{+\infty}dy {1\over {[P^+(y-x)]_{ML}}}{1\over {2(y-x)P^+
[k^--p^-]-({\bf k}-{\bf p})^2+i\epsilon}}=0.
\end{equation}
We stress that this identity holds only if the ML prescription is chosen.

By this subtraction the gauge pole is ``sterilized''; the integrations over
the minus components can be performed, now leading to the result
\begin{eqnarray}
\label{psiright}
&&\psi_{1GE}(x,{\bf p};s_1,s_2)=\theta(1-x) \theta(x)
\int {{d^2{\bf k}}\over {(2\pi )^3}}
\nonumber \\
&\times&\Big\{\int_{x}^{\infty} dy {\Big[P^+(y-x)\Big]}^{-1}
 \Big({{F(x,y,{\cal E}-\omega(1-y,{\bf k}^2)-\omega(x,{\bf p}^2),
{\bf k}, {\bf p})\theta(1-y)}\over {2(y-x)P^+[{\cal E}-\omega(1-y,{\bf k}^2)
-\omega(x,{\bf p}^2)]-({\bf k}-{\bf p})^2+i\epsilon}}
\nonumber \\
&-&{{F(x,x,{\cal E}-\omega(1-x,{\bf k}^2)-\omega(x,{\bf p}^2),
{\bf k}, {\bf p})}\over {2(y-x)P^+[{\cal E}-\omega(1-x,{\bf k}^2)
-\omega(x,{\bf p}^2)]-({\bf k}-{\bf p})^2+i\epsilon}}\Big)\nonumber \\
&+&\int_{-\infty}^{x} dy {\Big[P^+(y-x)\Big]}^{-1}
\Big({{F(x,y,-{\cal E}+\omega(y,{\bf k}^2)+\omega(1-x,{\bf p}^2),
{\bf k}, {\bf p})\theta(y)}\over {2(x-y)P^+[{\cal E}-\omega(y,{\bf k}^2)
-\omega(1-x,{\bf p}^2)]-({\bf k}-{\bf p})^2+i\epsilon}}
\nonumber \\
&-&{{F(x,x,-{\cal E}+\omega(x,{\bf k}^2)+\omega(1-x,{\bf p}^2),
{\bf k}, {\bf p})}\over {2(x-y)P^+[{\cal E}-\omega(x,{\bf k}^2)
-\omega(1-x,{\bf p}^2)]-({\bf k}-{\bf p})^2+i\epsilon}}\Big)\Big\}. 
\end{eqnarray} 
No end-point singularities are left after this procedure.

\subsection{The self-energy contributions}
\noindent
Now we turn our attention to the other three terms in eq. (\ref{Bsalpeter})
involving the self-energy in which we will retain, coherently with the
approximation done on the ``exchange'' graph, only the one-loop
contribution. One can have a self-energy insertion on the quark line,
on the antiquark line, or on both.

The terms involving the quark and the antiquark self-energy are, respectively
$$\Phi_{SE1}(p)=-iS(p)\Sigma_1(p)\Phi(p)$$ and $$\Phi_{SE2}(p)=
-i\Phi(p)\Sigma_1(p-P)S(p-P).$$
Here $\Sigma_1$ is the one-loop self-energy, renormalized in the
$\overline{{\rm MS}}$ scheme. The corresponding contributions to the null-plane
wave function are
\begin{equation}
\label{psise1}
\psi_{SE1}(x,{\bf p};s_1,s_2)=\int {{xdp^-}\over {2\pi}}{{\bar u(xP^+,
{\bf p};s_1)\Sigma_1(p)\Phi(p)\gamma^+v((1-x)P^+,-{\bf p};s_2)}\over
{p^2-m^2+i\epsilon}}
\end{equation}
and
\begin{equation}
\label{psise2}
\psi_{SE2}(x,{\bf p};s_1,s_2)=-\int {{(1-x)dp^-}\over {2\pi}}{{\bar u(xP^+,
{\bf p};s_1)\gamma^+\Phi(p)\Sigma_1(p-P)v((1-x)P^+,-{\bf p};s_2)}\over
{(p-P)^2-m^2+i\epsilon}},
\end{equation}
respectively.

In ref.\cite{SOPER93} it is carefully explained how the contribution from the
self-energy insertion on both 
quark and antiquark lines can be split into two pieces, one that will 
cancel part of $\Phi_{SE1}$ and another that will cancel part of $\Phi_{SE2}$.
These cancellations are part of the Tamm-Dancoff approximation we are
considering. We are thereby left with the following two self-energy
contributions:
\begin{eqnarray}
\label{self1}
&&\psi_{SE1}(x,{\bf p};s_1,s_2)={1\over {2xP^+}}\sum_{s_1'}\psi(x,{\bf p};
s_1',s_2)\nonumber \\
&\times& {{\bar u(xP^+,{\bf p};s_1)\Sigma_1(xP^+,{\cal E}-\omega(1-x,{\bf p}^2),
{\bf p})u(xP^+,{\bf p};s_1')}\over {{\cal E}-\omega(1-x,{\bf p}^2)-
\omega(x,{\bf p}^2)+i\epsilon}}
\end{eqnarray}
and
\begin{eqnarray}
\label{self2}
&&\psi_{SE2}(x,-{\bf p};s_1,s_2)=-{1\over {2(1-x)P^+}}\sum_{s_2'}
\psi(x,-{\bf p};
s_1,s_2')\nonumber \\
&\times& {{\bar v((1-x)P^+,-{\bf p};s_2')\Sigma_1(-(1-x)P^+,-{\cal E}+
\omega(x,{\bf p}^2),
{\bf p})v((1-x)P^+,-{\bf p};s_2)}\over {{\cal E}-\omega(1-x,{\bf p}^2)-
\omega(x,{\bf p}^2)+i\epsilon}}.
\end{eqnarray}

The unrenormalized quark self-energy in the one-loop approximation is given by
\begin{equation}
\label{sigma}
\Sigma_1^{\varepsilon}(p)=g^2C_F{\mu}^{2\varepsilon}
\int {{d^{4-2\varepsilon}k}\over 
{(2\pi)^{4-2\varepsilon}}}\gamma_{\alpha}S(k)\gamma_{\beta}
{{N^{\alpha\beta}(q)}
\over {[q\cdot n]_{ML}(q^2+i\epsilon)}},
\end{equation}
where $k^{\alpha}$ is the quark momentum and $q^{\alpha}
=p^{\alpha}-k^{\alpha}$
the gluon momentum. We use dimensional regularization, the coupling 
constant g is dimensionless and $\mu$ is the running
mass scale.

This equation can be rewritten as
\begin{equation}
\label{sigmad}
-i\Sigma_1^{\varepsilon}(p)=g^2C_F{\mu}^{2\varepsilon}\int {{d^{2-2
\varepsilon}{\bf k}}
\over 
{(2\pi)^{2-2\varepsilon}}}\Big[A+B^{\mu}\gamma_{\mu}+C^{\mu}\gamma_{\mu}\Big],
\end{equation}
where $B^{\mu}$ is the only term that receives a contribution from the gauge
pole. The quantities $A$, $B^{\mu}$ and $C^{\mu}$ 
are given by
\begin{equation}
\label{aterm}
A=\int {{dk^+dk^-}\over {(2\pi)^2}}
{{-2m(1-\e)}\over 
{(p-k)^2+i\epsilon}}{1\over{k^2-m^2+i\epsilon}},
\end{equation}
\begin{eqnarray}
\label{spuden}
B^{\mu}&=&\int {{dk^+dk^-}\over {(2\pi)^2}}{1\over {[p^+-k^+]_{ML}}}
{1\over {2(p^+-k^+)(p^--k^-)-({\bf p}-{\bf k})^2+i\epsilon}}
\nonumber \\
&\times&{{{\cal B}^{\mu}(k^+,k^-,{\bf k};p^+,p^-,{\bf p})}\over
{2k^+k^--{\bf k}^2 -m^2+i\epsilon}},
\end{eqnarray}
with
$${\cal B}^+=0,$$ $${\cal B}^-=4k^+(p^--k^-)-2{\bf k}\cdot({\bf p}-{\bf k}),$$
$${\cal B}^j=2k^+(p^j-k^j)-2(p^+-k^+)k^j $$
and
\begin{equation}
\label{cterm}
C^{\mu}=\int {{dk^+dk^-}\over {(2\pi)^2}}
{{2k^{\mu}(1-\e)}\over 
{(p-k)^2+i\epsilon}}{1\over{k^2-m^2+i\epsilon}}.
\end{equation}

The gauge singularity in eq. (\ref{spuden}), 
being prescribed according to (ML), does not spoil
the convergence of the integrals. In other words no
singularity of IR type occurs, thanks to the prescription, 
while UV singularities in eq. (\ref{sigmad})
are cured by dimensional regularization. In passing we stress that
this procedure has the merit of clearly disentangling possible IR and UV 
singularities.

Now the gauge pole can be ``sterilized'' by a suitable subtraction, 
in the same way as we did for the exchange term, thereby allowing us to
perform the integration over $k^-$; we obtain
\begin{eqnarray}
\label{spudent}
B^{\mu}&=&{{-i}\over {4\pi p^+}}\int_{-\infty}^{1}{{dx}\over {1-x}}
\Big\{{{\theta(x){\cal B^{\mu}}
(xp^+,{{{\bf k}^2+m^2}\over {2xp^+}},{\bf k};p^+,p^-,
{\bf p})}\over {2x(1-x)p^+p^--(1-x)({\bf k}^2+m^2)-x({\bf p}-{\bf k})^2
+i\epsilon}} \nonumber \\
&-&{{{\cal B^{\mu}}(p^+,{{{\bf k}^2+m^2}\over {2p^+}},{\bf k};p^+,p^-,
{\bf p})}\over {2(1-x)p^+p^--(1-x)({\bf k}^2+m^2)-({\bf p}-{\bf k})^2
+i\epsilon}}\Big\}.
\end{eqnarray}
We notice that the potential singularity at $x=1$ is cancelled by the
subtraction. We also notice that the two subtractions, the one in the
exchange term and the one in the self-energy expressions, 
although dictated
by a similar philosophy, have nothing to do with each other.

The null-plane wave function $\psi$ is eventually
obtained as
\begin{equation}
\label{sum}
\psi(x,{\bf p};s_1,s_2)=\psi_{1GE}(x,{\bf p};s_1,s_2)+
\psi_{SE1}(x,{\bf p};s_1,s_2)+\psi_{SE2}(x,{\bf p};s_1,s_2).
\end{equation}

At this point we have recalled from ref.\cite{SOPER93} all the concepts we need
to develop our argument in the next section.

\section{The Bethe-Salpeter equation in 1+1 dimensions}
\noindent

When going to 1+1 dimensions, UV singularities will no longer show
up; in turn the IR behaviour is worsened. Subtleties occur in 
this dimensional reduction.

We start from unrenormalized, dimensionally regularized quantities.
First, in 1+1 dimensions, the coupling constant acquires the 
dimension of a mass; this is automatically provided by the factor 
${\mu}^{2\e}$. But, in this case, the meaning of such a mass completely
changes: it is no longer a running mass scale, but rather it tunes the
dimensionful coupling, which is a free parameter 
characterizing the theory.

Second, the quantities $A$ and $C^{\mu}$ in eqs. (\ref{aterm}),(\ref{cterm})
vanish, in strictly 1+1 dimensions, as a consequence of the Dirac
algebra. However, if the calculation is performed in $4-2\e$
dimensions and the loop integration over transverse momenta is carried on,
the $\e -1$ zero coming from the polarization factor, is fully 
compensated by a pole, leading eventually in the limit $\e \to 1$
to the non vanishing expressions
\begin{equation}
\label{afinite}
-i\Sigma_A=-{{iC_Fg^2}\over{2\pi}}{m\over{p^2-m^2+i\epsilon}}
\end{equation}
and
\begin{equation}
\label{cfinite}
-i\Sigma_C=-{{iC_Fg^2}\over{2\pi}}{{p^{\mu}\gamma_{\mu}}
\over{p^2-m^2+i\epsilon}},
\end{equation}
where we have again denoted by $g$ the coupling constant of the theory,
which differs from the coupling constant of the previous Section 
by the factor $\mu$. 

We stress that the above quantities are sensitive to the way in which the 
transition to 1+1 dimensions is performed. This anomaly-type phenomenon
is reminiscent of an analogous effect we found in perturbative Wilson
loop calculations \cite{BASGRI} and is worthy of further study; it 
points towards a discontinuity of the theory in the limit $\varepsilon 
\to 1$ \cite{BASNAR}. 

The terms $B^{\mu}$, which are the ones affected by the gauge pole, are instead 
insensitive to the way in which
the reduction is performed: the same result is indeed obtained just 
ignoring transverse degrees of freedom or taking the limit $\e \to 1$ 
at the very end of the calculation. 

There is then the problem of formulating the gluon exchange contribution
in $4-2\varepsilon$ dimensions. To this purpose one should consider
unrenormalized quantities, which are expected to produce singularities
of UV nature just as poles at some integer values of dimensions.
Unfortunately, in order to decide whether the limit $\varepsilon \to 1$
is smooth, one should solve the integral equation for a generic 
value of $\varepsilon$, or, at least, to have a control on its
behavior with respect to transverse momentum.

We leave to a future investigation the interesting problem of studying the 
limit $\varepsilon \to 1$. In the sequel we adopt the attitude of
working directly in 1+1 dimensions, ``freezing'' the transverse degrees of
freedom.
We drop everywhere the transverse-momentum dependence in 
eq. (\ref{psirealapp}). This procedure turns the simple pole at $P^+(y-x)=0$
into a double pole. Integration over this double pole is perfectly
prescribed, though; thanks to the ML recipe, both singularities lie 
on the same side of the integration contour. In other words no pinch
occurs when dropping transverse momenta. Nevertheless a double pole
would require {\it two} subtractions to be sterilized.
We would like to stress again that this ``sterilization'' is {\it not}
required to give the integrals a meaning (they are indeed already 
perfectly defined), but motivated  by the desire to perform 
first the integration
over minus components in order to recover the null-plane
perturbative formulation.

We might operate subtractions also in this case, repeating the treatment 
of the previous section; however,
as it will become apparent that subtractions are not needed in 1+1
dimensions, we shall recover null-plane perturbation theory by following
a slightly different procedure.

In 1+1 dimensions, eq. (\ref{psirealapp}) becomes
\begin{eqnarray}
\label{psirealapp1}
\psi_{1GE}(x)&=&
-\int {{dy}\over {2\pi}} \int {{dk^-}\over {2\pi}} \int {{dp^-}\over {2\pi}} 
 F(x,y,k^--p^-)
\nonumber \\
&\times& \Big[{(k^--\omega(y)+i\epsilon \ {\rm sign}(y))}^{-1}-
{(k^--{\cal E}+\omega(1-y)-i\epsilon\ {\rm sign}(1-y))}^{-1}\Big]
\nonumber \\
&\times& \Big[{(p^--\omega(x)+i\epsilon\ {\rm sign}(x))}^{-1}-
{(p^--{\cal E}+\omega(1-x)-i\epsilon\ {\rm sign}(1-x))}^{-1}\Big]
\nonumber \\
&\times&{1\over {\Big[P^+(y-x)\Big]_{ML}}}{\Big[2P^+(y-x)(k^--p^-)
+i\epsilon\Big]}^{-1},
\end{eqnarray} 
with
\begin{eqnarray}
\label{bigeffe1}
&&F(x,y,k^--p^-)=-{{C_F g^2}\over {4P^+y(1-y)}}
\psi(y)\bar u(xP^+)
\gamma^+u(yP^+) \nonumber \\
&&{{2(k^--p^-)}\over {{\cal E}-\omega(1-x)-
\omega(x)+i\epsilon}}\bar v((1-y)P^+)\gamma^+
v((1-x)P^+).
\end{eqnarray} 
Taking the detailed expressions of the light-cone spinors into account
\cite{BRODSKY}, eq. (\ref{bigeffe1}) can be written as
\begin{equation}
\label{bigeffe2}
F(x,y,k^--p^-)=-2C_F g^2 P^+ \psi(y) {{\sqrt{x(1-x)}}\over {\sqrt{y(1-y)}}}
{{(k^--p^-)}\over {{\cal E}-\omega(1-x)-
\omega(x)+i\epsilon}}.
\end{equation} 
Equation (\ref{psirealapp1}) in turn becomes
\begin{eqnarray}
\label{psirealapp2}
\psi_{1GE}(x)&=&{{C_F g^2 P^+}\over {{\cal E}-\omega(1-x)-
\omega(x)+i\epsilon}} 
\int {{dy}\over {2\pi}} \int {{dk^-}\over {2\pi}} \int {{dp^-}\over {2\pi}} 
\psi(y) {{\sqrt{x(1-x)}}\over {\sqrt{y(1-y)}}}
\nonumber \\
&\times& \Big[{(k^--\omega(y)+i\epsilon \ {\rm sign}(y))}^{-1}-
{(k^--{\cal E}+\omega(1-y)-i\epsilon\ {\rm sign}(1-y))}^{-1}\Big]
\nonumber \\
&\times& \Big[{(p^--\omega(x)+i\epsilon\ {\rm sign}(x))}^{-1}-
{(p^--{\cal E}+\omega(1-x)-i\epsilon\ {\rm sign}(1-x))}^{-1}\Big]
\nonumber \\
&\times&{\Big[P^+(y-x)\Big]^{-2}_{ML}}.
\end{eqnarray} 

We have now reached a complete symmetry between gauge and ``Feynman'' pole.
This pole should be prescribed causally in equal-time quantization;
this is certainly mandatory when propagating transverse degrees of
freedom are present, {\it i.e.} in higher dimensions. Its causal prescription
forces the gauge pole to be causal too, for consistency. On the other hand,
the causal option follows from equal-time quantization \cite{BASGRI}.

Let us now go back to eq. (\ref{mandel}) and consider the identity
\begin{equation}
\label{idman}
{1\over {[q^+]_{ML}}}
\equiv {1\over {q^++i\epsilon \ {\rm sign}(q^-)}} =
{1\over {[q^+]_{CPV}}}-i\pi \ {\rm sign}(q^-)\delta (q^+),
\end{equation} 
which, after differentiation with respect to $q^+$, becomes
\begin{equation}
\label{diffid}
{1\over {[q^+]^2_{ML}}}={1\over {[q^+]^2_{CPV}}}+i\pi \ {\rm sign}(q^-)
\delta^{\prime} (q^+).
\end{equation}
At this point it is convenient to change the normalization of the function 
$\psi$, by defining $$\psi(x)=\phi(x) \sqrt {x(1-x)}.$$
Introducing eq. (\ref{diffid}) in eq. (\ref{psirealapp2}), we obtain
\begin{equation}
\label{psirealsplit}
\phi_{1GE}(x)=\ \phi^{CPV}_{1GE}(x)+\ \phi^{(s)}_{1GE},
\end{equation}
with
\begin{eqnarray}
\label{psicpv}
\phi^{CPV}_{1GE}(x)&=&{{C_F g^2 P^+}\over {{\cal E}-\omega(1-x)-
\omega(x)+i\epsilon}} 
\int {{dy}\over {2\pi}} \int {{dk^-}\over {2\pi}} \int {{dp^-}\over {2\pi}} 
\phi(y) 
\nonumber \\
&\times& \Big[{(k^--\omega(y)+i\epsilon \ {\rm sign}(y))}^{-1}-
{(k^--{\cal E}+\omega(1-y)-i\epsilon\ {\rm sign}(1-y))}^{-1}\Big]
\nonumber \\
&\times& \Big[{(p^--\omega(x)+i\epsilon\ {\rm sign}(x))}^{-1}-
{(p^--{\cal E}+\omega(1-x)-i\epsilon\ {\rm sign}(1-x))}^{-1}\Big]
\nonumber \\
&\times&{\Big[P^+(y-x)\Big]^{-2}_{CPV}}
\end{eqnarray} 
and
\begin{eqnarray}
\label{psisub}
\phi^{(s)}_{1GE}(x)&=&{{i\pi C_F g^2 P^+}\over {{\cal E}-\omega(1-x)-
\omega(x)+i\epsilon}} 
\int {{dy}\over {2\pi}} \int {{dk^-}\over {2\pi}} \int {{dp^-}\over {2\pi}} 
\phi(y) 
\nonumber \\
&\times& \Big[{(k^--\omega(y)+i\epsilon \ {\rm sign}(y))}^{-1}-
{(k^--{\cal E}+\omega(1-y)-i\epsilon\ {\rm sign}(1-y))}^{-1}\Big]
\nonumber \\
&\times& \Big[{(p^--\omega(x)+i\epsilon\ {\rm sign}(x))}^{-1}-
{(p^--{\cal E}+\omega(1-x)-i\epsilon\ {\rm sign}(1-x))}^{-1}\Big]
\nonumber \\
&\times&\ {\rm sign}(k^--p^-)\delta^{\prime}(P^+(y-x))\ .
\end{eqnarray}
In eq. (\ref{psicpv}) the integrations over the minus components of the momenta
can be
easily performed, leading to the expression
\begin{equation}
\label{psicpvi}
\phi^{CPV}_{1GE}(x)=-{{C_F g^2}\over { P^+[{\cal E}-\omega(1-x)-
\omega(x)+i\epsilon]}} 
\int_{0}^{1} {{dy}\over {2\pi}} \phi(y) 
{\Big[(y-x)\Big]}^{-2}_{CPV}\ .
\end{equation} 
In turn, eq. (\ref{psisub}) becomes
\begin{eqnarray}
\label{psisubsplit}
&&\phi^{(s)}_{1GE}(x)=-{{i\pi C_F g^2 }\over {{\cal E}-\omega(1-x)-
\omega(x)+i\epsilon}} 
\int {{dy}\over {2\pi}} \int {{dk^-}\over {2\pi}} \int {{dp^-}\over {2\pi}} 
\delta(y-x) \ {\rm sign}(k^--p^-) 
\nonumber \\
&\times& \Big[{(p^--\omega(x)+i\epsilon\ {\rm sign}(x))}^{-1}-
{(p^--{\cal E}+\omega(1-x)-i\epsilon\ {\rm sign}(1-x))}^{-1}\Big]
\nonumber \\
&\times&\Big(\ \phi^{\prime}(y) \Big[{(k^--\omega(y)+i\epsilon\ {\rm sign}(y))}^{-1}-
{(k^--{\cal E}+\omega(1-y)-i\epsilon\ {\rm sign}(1-y))}^{-1}\Big]\nonumber \\
&+&\phi(y) {d\over {dy}} \Big[{(k^--\omega(y)+i\epsilon\ {\rm sign}(y))}^{-1}-
{(k^--{\cal E}+\omega(1-y)-i\epsilon\ {\rm sign}(1-y))}^{-1}\Big]\Big)\ .
\end{eqnarray}
Now integrations over the minus components can be done; the first
term vanishes for symmetry reasons; the second one, after some algebra, 
taking the expression for $\omega$ into account,
becomes
\begin{equation}
\label{psisubfin}
\phi^{(s)}_{1GE}(x)={{g^2 m^2 C_F \phi(x)}\over {4\pi {(P^+)}^2}}
{1\over {({\cal E}-\omega(x)-\omega(1-x)+i\epsilon)^2}}\ [x^{-2}+
(1-x)^{-2}].
\end{equation}

\bigskip

Then we repeat the treatment in the expressions concerning the self-energy
contributions.
Let us
therefore go back to eq. (\ref{spuden}), which, in 1+1 dimensions,
becomes
\begin{equation}
\label{spuden2}
B^-=\int {{2 k^+dk^+dk^-}\over {(2\pi)^2}}{1\over {[p^+-k^+]_{ML}^2}}
{1 \over
{2k^+k^-m^2+i\epsilon}}={{ip^-}\over {\pi (p^2-m^2+i\epsilon)}}\  .
\end{equation}
Using the identity
(\ref{diffid}), we obtain the splitting 
\begin{equation}
\label{spuden2split}
B^-={i\over {2\pi p^+}}+{{im^2}\over {\pi (2xP^+)^2}}
{1\over {{\cal E}-\omega(x)-\omega(1-x)+i\epsilon}}
\end{equation}
and, correspondingly,
\begin{equation}
\label{self12}
\phi_{SE1}(x)=\phi^{CPV}_{SE1}(x)+\phi^{(s)}_{SE1}(x)
\end{equation}
with
\begin{equation}
\label{self1cpv}
\phi^{CPV}_{SE1}(x)=-
{{g^2 C_F}\over {2\pi xP^+}}\phi(x)
{1\over {{\cal E}-\omega(1-x)-
\omega(x)+i\epsilon}}
\end{equation}
and
\begin{equation}
\label{self1s}
\phi^{(s)}_{SE1}(x)=-
{{g^2 m^2 C_F}\over {\pi (2xP^+)^2}}\phi(x)
{1\over {({\cal E}-\omega(1-x)-
\omega(x)+i\epsilon)^2}}\ .
\end{equation}
Similarly, for the second self-energy contribution we get
\begin{equation}
\label{selffi}
\phi_{SE2}(x)=\phi^{CPV}_{SE2}(x)+\phi^{(s)}_{SE2}(x),
\end{equation}
with
\begin{equation}
\label{self2cpv}
\phi^{CPV}_{SE2}(x)=-
{{g^2 C_F}\over {2\pi (1-x)P^+}}\phi(x)
{1\over {{\cal E}-\omega(1-x)-
\omega(x)+i\epsilon}}
\end{equation}
and
\begin{equation}
\label{self2s}
\phi^{(s)}_{SE2}(x)=-
{{g^2 m^2 C_F}\over {\pi (2(1-x)P^+)^2}}\phi(x)
{1\over {({\cal E}-\omega(1-x)-
\omega(x)+i\epsilon)^2}}\ .
\end{equation}

\bigskip

Summing everything together, we find that all $\phi^{(s)}$'s cancel
and we are left with:
\begin{eqnarray}
\label{psifinal}
&&\phi(x)=-{{C_F g^2}\over {2\pi P^+[{\cal E}-\omega(1-x)-
\omega(x)+i\epsilon]}} 
\nonumber \\
&\times&\Big[{{\phi(x)}\over {x(1-x)}}+ \int_{0}^{1} 
dy \phi(y)
{\Big[(y-x)\Big]}^{-2}_{CPV}\Big]\ .
\end{eqnarray} 
ML and CPV are completely equivalent in this case!

We remark that eq. (\ref{psifinal}) is nothing but 't Hooft's
equation \cite{THOOFT}, in spite of the seemingly different physical 
inputs. 

\section{Final remarks}
\noindent
We started by considering a ``causal'' formulation of the 
bound-state integral equation in the lowest-order
Tamm-Dancoff approximation, in particular by considering only one-loop
contributions to the self-energy, and then, after a suitable
dimensional reduction, we ended up with 't Hooft's
equation in which all planar diagrams are summed (large-$N$ approximation)
with an ``instantaneous'' potential between quarks. How did it
happen?

The reason why ``causal'' and ``instantaneous'' interactions 
lead to the
same answer in this case has already been anticipated; it is rooted in
the cancellations occurring in 1+1 dimensions thanks to one-loop unitarity.
Those cancellations had already been noticed \cite{BASGRIG}, although
in a different context and with a different technique.

In turn the reason why the Tamm-Dancoff
approximation reproduces 't Hooft's
full planar summation is due to the dynamical circumstance
that in 't Hooft's formulation the exact solution for the self-energy
coincides with its ${\cal O}(g^2)$ expression. 

This fact also explains why we did not recover
Wu's equation, when considering the ``causal'' formulation. As a matter
of fact, in Wu's treatment
the exact solution for the
self-energy exhibits a quite involved analytical structure;
in particular, it does not generally match, in the relevant Ward identity,
the expression used for the vertex in the bound-state equation. 

Since at large $N$ the same set of diagrams, the planar ones, 
are summed in both formulations, 
we envisage a potential conflict, beyond the one-loop approximation,
between planarity and ``causal''
formulation in 1+1 dimensions. 
This crucial issue in our opinion deserves further study.

In dimensions higher than 2, causality looks mandatory
and only one formulation (the ``causal'' one) can reasonably survive.

\vfill\eject

\end{document}